\documentclass[12pt]{article}
\usepackage{epsf}
\def\be{\begin{equation}}
\def\ee{\end{equation}}
\def\bea{\begin{eqnarray}}
\def\eea{\end{eqnarray}}

\begin{document}
\begin{titlepage}
\begin{center}
{\Large \bf William I. Fine Theoretical Physics Institute \\
University of Minnesota \\}
\end{center}
\vspace{0.2in}
\begin{flushright}
FTPI-MINN-14/13 \\
UMN-TH-3336/14 \\
May 2014 \\
\end{flushright}
\vspace{0.3in}
\begin{center}
{\Large Contribution of $Z_b$ resonances to $\Upsilon(5S) \to \pi \pi \pi \chi_b$.
\\}
\vspace{0.2in}
{\bf Xin Li$^a$  and M.B. Voloshin$^{a,b,c}$  \\ }
$^a$School of Physics and Astronomy, University of Minnesota, Minneapolis, MN 55455, USA \\
$^b$William I. Fine Theoretical Physics Institute, University of
Minnesota,\\ Minneapolis, MN 55455, USA \\
$^c$Institute of Theoretical and Experimental Physics, Moscow, 117218, Russia
\\[0.2in]

\end{center}

\vspace{0.2in}

\begin{abstract}
We discuss the recently presented Belle results on the decays $\Upsilon(5S) \to \pi \pi \pi \chi_{bJ}(1P)$. The data indicate that in addition to the $\omega$ emission, $\Upsilon(5S) \to \omega \chi_{bJ}$, there is a significant non resonant background in the three pion spectrum. We suggest that a sizable fraction of this background may be associated with the cascade process $\Upsilon(5S) \to \pi Z_b \to \pi \rho \chi_b$ involving the $Z_b(10610)$ and $Z_b(10650)$ resonances. If confirmed by the data, this would be the first observation of transition from the $Z_b$ resonances to lower bottomonium with emission of a light meson state different from a single pion, which may provide a new input in understanding of the internal dynamics of these resonances.
\end{abstract}
\end{titlepage}

The recently reported\,\cite{Shen} preliminary results of analysis of the Belle data on the decays of the $\Upsilon(5S)$ [$\Upsilon(10860)$] bottomonium resonance to $\pi^+ \pi^- \pi^0 \chi_{bJ}$ with $J=1,2$\,\footnote{The contribution of the $J=0$ bottomonium state, $\chi_{b0}$, is suppressed due to the requirement in the analysis that the final $\chi_{bJ}$ state radiative decay into $\gamma + \Upsilon(1S)$ is also observed.} present an observation of the hadronic transitions with emission of the $\omega$ resonance: $\Upsilon(5S) \to \omega \chi_{bJ}$. In addition, the data also reveal a significant non-$\omega$ background in the invariant mass distribution of the three pions, which distribution is enhanced at the higher end of the spectrum around $0.9\,$GeV. We suggest here that the non-$\omega$ part of the process may be in fact due to the contribution of the bottomonium-like isovector $Z_b$ [$Z_b(10610)$] and $Z_b'$ [$Z_b(10650)$] resonances through the cascade transitions $\Upsilon(5S) \to \pi Z_b^{(')} \to \pi \rho \chi_{bJ}$ shown in Fig.1, naturally resulting in the enhancement in the three-pion invariant mass spectrum around $0.9\,$GeV, although the presence of the $Z_b$ resonances can be better studied e.g. by the energy distribution of a single pion. In particular, the rate for the non-$\omega$ background suggests the possibility that the couplings of the $Z_b$ and $Z_b$ resonances to the channels $\pi h_b$ and $\rho \chi_{bJ}$ are very similar, which might hint at yet unknown relations in the dynamics of light quarks in the presence of a heavy quark-antiquark pair.

\begin{figure}[ht]
\begin{center}
 \leavevmode
    \epsfxsize=10cm
    \epsfbox{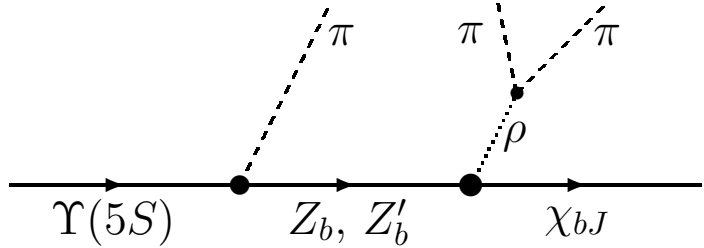}
    \caption{The cascade process $\Upsilon(5S) \to \pi Z_b^{(')} \to \pi \rho \chi_{bJ}$ in the decay $\Upsilon(5S) \to \pi \pi \pi \chi_{bJ}(1P)$. }
\end{center}
\end{figure}

The $Z_b^{(')}$ resonances were found\,\cite{bellez} by their decays to lower bottomonium states with emission of a single pion: $Z_b^{(')} \to \pi \Upsilon(nS)$ with $n=1,2,3$ and $Z_b^{(')} \to \pi h_b(kP)$ with $k=1,2$. These states are interpreted\,\cite{bgmmv} as `molecular' $S$-wave threshold resonances made of heavy bottom meson-antimeson pairs: $Z_b \sim (B \bar B^* - \bar B B^*)$, $Z_b' \sim B^* \bar B^*$, with the quantum numbers $I^G(J^P)=1^+(1^+)$. In such molecular states the spins of the heavy quark and antiquark are not correlated with each other, but rather with the spin of the light (anti)quark in the heavy meson. As a result, the resonances $Z_b$ and $Z_b'$ are (orthogonal) mixed states with respect to the total spin of the $b \bar b$ quark pair:
\be
Z_b \sim  1^-_{b \bar b} \otimes 0^-_L + 0^-_{b \bar b} \otimes 1^-_L ~,~~~~~  Z_b' \sim  1^-_{b \bar b} \otimes 0^-_L - 0^-_{b \bar b} \otimes 1^-_L ~, 
\label{zbs}
\ee
where $J^P_{b \bar b}$ stand for the total spin of the heavy $b \bar b$ quark pair, and $J^P_L$ is the same for the rest (light) degrees of freedom. In particular, this structure describes well\,\cite{bgmmv,xlmv} the comparable decay rate of the $Z_b^{(')}$ resonances to spin triplet [$\Upsilon(nS)$] and spin singlet [$h_b(kP)$] lower bottomonium states. 

The quantum numbers and the heavy quark spin structure of the $Z_b^{(')}$ states suggest that besides the observed single pion transitions to the lower bottomonium states there should exist\,\cite{bgmmv} similar transitions with emission of two pions: $Z_b^{(')} \to \rho \eta_b(1S) \to \pi \pi \eta_b(1S)$ and $Z_b^{(')} \to \rho \chi_{bJ}(1P) \to \pi \pi \chi_{bJ}(1P)$. In most of the latter transitions to the $\chi_{bJ}(1P)$ levels the energy is slightly below the nominal mass of the $\rho$ meson [except for the transition $Z_b' \to \rho \chi_{b0}(1P)$], but the resulting kinematical suppression is not very strong due to large width of the $\rho$ resonance. Clearly, the latter transitions would contribute to the observed signal for the decay $\Upsilon(5S) \to \pi \pi \pi \chi_{bJ}(1P)$ as a non-$\omega$ background. 

In the limit of heavy quark spin symmetry (HQSS) all six transitions from both $Z_b$ and $Z_b'$ resonances to the three $\chi_{bJ}$ levels proceed due to the part of the spin wave function (\ref{zbs}) containing the $1^-_{b \bar b}$ component, and their  amplitudes $A(Z_b^{(')} \to \rho \chi_{bJ})$ can be described by one coupling $g_\rho$:
\be
A(Z_b^{(')} \to \rho \chi_{bJ}) = g_\rho \, \left ( Z_i^a+{Z_i^a}' \right ) \, q_j \, \rho_k^a \, \left [ {1 \over \sqrt{3} } \epsilon_{ijk} \, \chi^{(0)} + {1 \over \sqrt{2}} \, \left ( \delta_{ij} \, \chi^{(1)}_k - \delta_{ik} \, \chi^{(1)}_j \right ) + \epsilon_{jkl} \, \chi^{(2)}_{il} \right ]~,
\label{amp6}
\ee
where $a$ is the isotopic triplet index, $\vec q$ is the momentum of the $\rho$ meson, $\vec Z$, $\vec Z'$ and $\vec \rho$ are the polarization amplitudes of respectively the $Z_b$, $Z_b'$ and $\rho$ resonances, and $\chi^{(0)}$, $\chi^{(1)}_i$ and $\chi^{(2)}_{ij}$ stand for the amplitudes of the final $\chi_{bJ}$ states with respectively $J=0,1$ and $2$. The latter amplitudes are assumed to be normalized to the number of polarization states: $\chi^{(0)} \chi^{(0)*}=1$, $\chi^{(1)}_i \chi^{(1)*}_i=3$, $\chi^{(2)}_{ij} \chi^{(2)*}_{ij} =5$, and the spin-2 amplitude $\chi^{(2)}_{ij}$ is symmetric and traceless.

It can be noted that the equal coupling of the $Z_b$ and $Z_b'$ resonances in Eq.(\ref{amp6}) is also a consequence of HQSS, since these resonances become degenerate in the limit where the interaction due to the heavy quark spin is turned off\,\cite{bgmmv}. This assumption is known to be in a reasonable agreement (within the current experimental uncertainty) with the available data\,\cite{bellez} on the relative strength of the $Z_b$ and $Z_b'$ peaks in the channels $\pi \Upsilon(nS)$ and $\pi h_b(kP)$. 

We also emphasize here that an application of HQSS to the discussed transitions from the  $Z_b^{(')}$ states has a quite different status than in the case of the decays from $\Upsilon(5S)$, where, e.g. in the decays $\Upsilon(5S) \to \omega \chi_{bJ}$, the measured\,\cite{Shen} yield of $\chi_{b1}$ is approximately three times larger than of $\chi_{b2}$, while a straightforward application of HQSS and treating $\Upsilon(5S)$ as a pure $b \bar b$ quarkonium would imply that relative yield should be 3:5 (i.e. proportional to the number of spin states for $\chi_{bJ}$), modulo minor kinematical corrections. However, there are reasons to conclude\,\cite{mv12} that the resonance $\Upsilon(5S)$ is a more complicated object, likely due to a mixing with states of heavy meson pairs, whose thresholds are close to its mass. In particular the reported\,\cite{bellez} presence of the $f_2(1270)$ tensor resonance in the dipion channel in the decay $\Upsilon(5S) \to \pi \pi \Upsilon(1S)$ is likely an indicator of the presence of in the $\Upsilon(5S)$ of a state with light degrees of freedom, denoted in Ref.\,\cite{mv12} as $\psi_{12}$, where the heavy $b \bar b$ pair has total spin one, while the light degrees of freedom are in a $J^{PC}=2^{++}$ state. For the decay of such state to $\omega \chi_{bJ}$, the HQSS would predict the ratio of the rates for $\chi_{b0} : \chi_{b1} :\chi_{b2} = 20:15:1$. Clearly, the experimentally observed\,\cite{Shen} ratio of $\chi_{b1} : \chi_{b2}$ can well be a result of a combined effect of the pure $b \bar b$ and $\psi_{12}$ components of the $\Upsilon(5S)$. The situation is different for the $Z_b^{(')}$ resonances. Their quantum numbers limit the possible internal spin structures to only those included in Eq.(\ref{zbs}) and it is only the part containing $1^-_{b \bar b}$ that, according to HQSS, gives rise to transitions to the spin-triplet quarkonium levels, including the decays to $\rho \chi_{bJ}$. It can also be mentioned that this spin structure of the $Z_b^{(')}$ resonances requires the statistical weight enhancement 5:3 for the yield of the $\chi_{b2}$ state relative to that of the $\chi_{b1}$ which appears to agree with the data indicating a somewhat larger non-$\omega$ background in the decays to $J=2$ state as compared to $J=1$, i.e. in a strong variance with the ratio for the resonant $\omega$ yield. We defer a more specific discussion of this point until our further presentation of numerical estimates.

The contribution of the $Z_b^{(')}$ resonances to the decays of the $\Upsilon(5S)$ depend on the coupling $f$ in the amplitude of the decay $\Upsilon(5S) \to \pi Z_b $, which amplitude has the form
\be
A[\Upsilon(5S) \to \pi Z_b^{(')} ] = f \, \left [ {\vec \Upsilon} \cdot \left ( \vec Z_b^a + \vec {Z_b^a}' \right ) \right ] \, \pi^a \,  E_\pi~, 
\label{ayz}
\ee
where the factor of the pion energy, $E_\pi$, is mandated by the chiral properties of soft pions\,\cite{bgmmv}.
In order to avoid the uncertainty related to the absolute value of this coupling $f$, we consider the ratio of the rate for the cascade decay  $\Upsilon(5S) \to \pi Z_b^{(')} \to \pi \rho \chi_{bJ}$ to the known one for the process  $\Upsilon(5S) \to \pi Z_b^{(')} \to \pi \pi h_b(1P)$, in which ratio the coupling $f$ cancels. Instead, this ratio depends on the relation between the constant $g_\rho$ in Eq.(\ref{amp6}) and a similar constant $g_\pi$ describing the transitions $Z_b^{(')} \to \pi h_b(1P)$ whose amplitude can be written as
\be
A[Z_b^{(')} \to \pi h_b(1P)]= g_\pi \, \epsilon_{ijk} \,  \left ( Z_i^a-{Z_i^a}' \right ) \, p_j \, h_k \pi^a~,
\label{amp2}
\ee
where $\vec h$ is the polarization amplitude of the $h_b(1P)$ bottomonium and $\vec p$ stands for the pion momentum. [It can be noticed that the relative sign between the amplitudes for the $Z_b$ and $Z_b'$ resonances in the transitions involving spin-singlet $b \bar b$ pair in Eq.(\ref{amp2}) is opposite to that for the transitions between spin-triplet states as in the amplitudes (\ref{amp6}) and (\ref{ayz}).]

Moreover, it is the ratio of the constants $g_\rho/g_\pi$ that can be of a particular interest. Indeed, in the HQSS limit the spin-singlet bottomonium $h_b(1P)$ and the spin-triplet states $\chi_{bJ}(1P)$ are described by the same spatial wave function. Therefore this ratio of the constants is determined by the relation between the wave functions for the triplet and the singlet in the spin of the $b \bar b$ pair parts of the $Z_b^{(')}$ resonances in Eq.(\ref{zbs}) and the relation between the amplitudes for conversion of the $0^-_L$ and $1^-_L$ light quark states into respectively  $\rho$ or a pion in the discussed hadronic transitions. (The change in the angular momentum is obviously compensated by the excitation of the $P$ wave in the final bottomonium and the $P$ wave motion of the light meson.) The numerical estimates, to be discussed further, suggest a tantalizing possibility of the constants $g_\rho$ and $g_\pi$ being (approximately) equal. At present we can offer no apriori motivation for such an equality to hold, however should it be established experimentally, it may indicate some kind of (approximate) symmetry in the dynamics of the molecular states.   

For the purpose of our numerical estimates we treat the $\rho$ meson as a simple Breit-Wigner resonance with fixed width $\Gamma_\rho \approx 150\,$MeV, thus neglecting the variation of its width parameter at the invariant mass of the $\pi \pi$ pair, $q^2$, being not equal to the nominal value of $m_\rho^2$. As is well known, such variation is process dependent, and, if required, can be studied and taken into account when (and if) more detailed data on the process $\Upsilon(5S) \to \pi^+ \pi^- \pi^0 \chi_{bJ}$ become available. We also note that the three terms arising in the amplitude from the isotopic permutation of the pion emerging from the first transition in the cascade, $\Upsilon(5S) \to \pi  Z_b^{(')}$, with one of the pions emerging from the $\rho$ resonance do not interfere in the total rate. This is due to that the pion in the first transition is emitted in the $S$-wave in the rest frame of the bottomonium [cf. Eq.(\ref{ayz})], while each of the pions from the decay of the $\rho$ resonance is in the $P$ wave in this frame, when the $\rho$ emission is described by the amplitude in Eq.(\ref{amp6}). After these preliminary remarks we write the expression for the ratio of the rates for the processes $\Upsilon(5S) \to \pi  Z_b^{(')} \to \pi^+ \pi^- \pi^0 \chi_{bJ}$ and  $\Upsilon(5S) \to \pi  Z_b^{(')} \to \pi^+ \pi^- h_b$ as follows
\be
{\Gamma [ \Upsilon(5S) \to \pi  Z_b^{(')} \to \pi^+ \pi^- \pi^0 \chi_{bJ}] \over \Gamma [ \Upsilon(5S) \to \pi  Z_b^{(')} \to \pi^+ \pi^- h_b ]} = {2 J+1 \over 2} \, {|g_\rho|^2 \over |g_\pi|^2 } \, {I_\rho \over I_\pi}~,
\label{rr}
\ee
where the phase space integrals $I_\rho$ and $I_\pi$ are given by
\be
I_\rho = \int \, {d E_\pi \, d q^2 \over \pi} \, \left | {1 \over E_1-E_\pi + i \Gamma_1/2} +  {1 \over E_2-E_\pi + i \Gamma_2/2} \right |^2 \, E_\pi^2 \, |\vec k_\pi| \, |\vec q|^3 \, { m_\rho \, \Gamma_\rho \over (q^2 - m_\rho^2)^2 + m_\rho^2 \, \Gamma_\rho^2}
\label{irho}
\ee
and
\be
I_\pi=\int \, d E_\pi \, \left | {1 \over E_1-E_\pi + i \Gamma_1/2} -  {1 \over E_2-E_\pi + i \Gamma_2/2} \right |^2 \, E_\pi^2 \, |\vec k_\pi| \, |\vec p|^3~.
\label{ipi}
\ee
Here the following notations are used: $\Gamma_1$ and $\Gamma_2$ are the widths of the $Z_b$ and $Z_b'$ resonances, $E_1$ and $E_2$ are the corresponding resonance energies for the pion emitted in the $\Upsilon(5S) \to \pi  Z_b^{(')}$ transition, $E_1 \approx 258\,$MeV, $E_2 \approx 213\,$MeV, $E_\pi$ is the energy of this pion, and $\vec k_\pi$ is its momentum. Furthermore, $q^2$ is the squared invariant mass of the two pions emerging from the $\rho$ decay, and $\vec q$ is the spatial part of $q$, $|\vec q|^2 = (\Delta - E_\pi)^2 - q^2$, with $\Delta$ being the total energy release in the transition from the initial $\Upsilon(5S)$ to the final $1P$ state of bottomonium. Finally $\vec p$ in Eq.(\ref{ipi}) is the momentum of the `second' pion, i.e. of the one emitted in the transition $Z_b^{(')} \to \pi h_b$. The kinematical boundaries in the integral (\ref{irho}) are $4 m_\pi^2 < q^2 < (\Delta - E_\pi)^2$ for the (first) integration over $q^2$ and $m_\pi < E_\pi < (\Delta - 2 m_\pi)$ for $E_\pi$, while in the integral $I_\pi$ the integration limits are set by $m_\pi < E_\pi < (\Delta - m_\pi)$ (although in practice both integrals are dominated by the values of $E_\pi$ in the vicinity of the resonances at $E_1$ and $E_2$). It can be also pointed out that the overall numerical factor in Eq.(\ref{rr}) corresponds to the normalization to the rate of the dipion transition to $h_b(1P)$ with only charged pions, for which rate the data are available\,\cite{bellez,pdg}\,\footnote{The simple model for the $\rho$ peak with $q^2$ independent width, which we use in our estimates, can be readily modified by replacing in Eq.(\ref{irho}) the constant $\Gamma_\rho$ with $\Gamma_\rho(q^2)$.}. Moreover, unlike the dipion transitions from $\Upsilon(5S)$ to $\Upsilon(nS)$ levels, which contain a nonresonant background not associated with the $Z_b$ and $Z_b'$ resonances, the transitions to the $h_b$ states are exclusively given by the resonance contribution, which is theoretically justified by the notion that the $Z_b^{(')}$ resonances provide the only significant mechanism for the apparent HQSS breaking and which behavior is in a very good agreement with the data\,\cite{bellez}. Thus the relation (\ref{rr}) applies to only the resonant process $\Upsilon(5S) \to \pi Z_b^{(')} \to \pi \pi \pi \chi_{bJ}$ rather than to any additional nonresonant background that may be present in $\Upsilon(5S)  \to \pi \pi \pi \chi_{bJ}$.  

Using the value 10.865\,GeV for the central energy at which the $\Upsilon(5S)$ data are collected by Belle, and also the current experimental central values of the widths for the $Z_b^{(')}$ resonances, $\Gamma_1=18.6\,$MeV and $\Gamma_2=11.5\,$MeV, we estimate numerically: $I_\pi \approx 2.94\,$GeV$^5$, $I_\rho(\chi_{b0}) \approx 0.231\,$GeV$^5$, $I_\rho(\chi_{b1}) \approx 0.152\,$GeV$^5$ and $I_\rho(\chi_{b2}) \approx 0.120\,$GeV$^5$. Therefore the relative yield of $\chi_{b0}:\chi_{b1}:\chi_{b2}$ bottomonium states in the discussed cascade process is estimated as
\be
1:2.0:2.6~.
\label{ry}
\ee
In terms of the ratio of the rate to that of $\Upsilon(5S) \to \pi^+ \pi^- h_b(1P)$ we find
\bea
&&{\Gamma [ \Upsilon(5S) \to \pi  Z_b^{(')} \to \pi^+ \pi^- \pi^0 \chi_{b1}] \over \Gamma [ \Upsilon(5S) \to \pi  Z_b^{(')} \to \pi^+ \pi^- h_b ]} \approx 0.078 \, {|g_\rho|^2 \over |g_\pi|^2 }~ , \nonumber \\ 
&&{\Gamma [ \Upsilon(5S) \to \pi  Z_b^{(')} \to \pi^+ \pi^- \pi^0 \chi_{b2}] \over \Gamma [ \Upsilon(5S) \to \pi  Z_b^{(')} \to \pi^+ \pi^- h_b ]} \approx 0.10 \, {|g_\rho|^2 \over |g_\pi|^2 }~.
\label{nrr}
\eea
The branching fraction for the decay $\Upsilon(5S) \to \pi^+ \pi^- h_b$ is currently measured\,\cite{pdg} as $ \left ( 3.5 ^{+1.0}_{-1.3} \right ) \times 10^{-3}$ which, using the estimates (\ref{nrr}), results in the predictions 
\bea
&&{\cal B} [ \Upsilon(5S) \to \pi  Z_b^{(')} \to \pi^+ \pi^- \pi^0 \, \chi_{b1}] \approx \left ( 2.7 ^{+0.8}_{-1.0} \right ) \times 10^{-4} {|g_\rho|^2 \over |g_\pi|^2 }~ , \nonumber \\
&&{\cal B} [ \Upsilon(5S) \to \pi  Z_b^{(')} \to \pi^+ \pi^- \pi^0 \, \chi_{b2}] \approx \left ( 3.5 ^{+1.0}_{-1.3} \right ) \times 10^{-4} {|g_\rho|^2 \over |g_\pi|^2 }~.
\label{nbr}
\eea
Experimentally the non-$\omega$ background in the Belle data\,\cite{Shen} on the decays $\Upsilon(5S) \to \pi^+ \pi^- \pi^0 \, \chi_{bJ}$ corresponds to the branching fraction of about $4 \times 10^{-4}$ for $\chi_{b1}$ and $7 \times 10^{-4}$ for $\chi_{b2}$ with an error in each value apparently amounting to between 2 and 3 units times $10^{-4}$.  

The non-$\omega$ background in the decays $\Upsilon(5S) \to \pi \pi \pi \chi_{bJ}(1P)$ can be only partially contributed by the $Z_b$ and $Z_b'$ resonances, similarly to only partial contribution of these resonances to decays $\Upsilon(5S) \to \pi \pi \Upsilon(nS)$. However, even given the current large uncertainty, the data\,\cite{Shen} suggest the possibility that the contribution of the $Z_b$ and $Z_b'$ resonances can be close to what one would estimate from Eq.(\ref{nrr}) with $|g_\rho| \approx |g_\pi|$.  Such relation could imply a spin independence not only for the heavy quarks, but also (an approximate one) for the light ones, in the processes of conversion of molecular states into a light meson and heavy quarkonium. In view of this intriguing possibility it appears very interesting to study these processes in more detail. In particular, the contribution of the $Z_b^{(')}$ states to the decays $\Upsilon(5S) \to \pi \pi \pi \chi_{bJ}(1P)$ can be evaluated from the data by the distribution of the smallest of the energies of the three pions in the decay, which distribution should contain the resonance peaks at $E_1$ and $E_2$. A relation between the emission of the pion and $\rho$ from different components of the $Z_b^{(')}$ resonances could possibly be also studied if the process $\Upsilon(5S) \to \pi Z_b^{(')} \to \pi \rho \eta_b(1S)$ was observed. Indeed, in this process the final bottomonium state $\eta_b(1S)$ is related by HQSS to the final state in the known process $\Upsilon(5S) \to \pi Z_b^{(')} \to \pi \pi \Upsilon(1S)$, and the relation between the couplings in the amplitudes of $Z_b^{(')} \to \rho \eta_b(1S)$ and $Z_b^{(')} \to \pi \Upsilon(1S)$ could thus be investigated.

We thank A.~Bondar for a useful discussion of the experimental possibilities. This work is supported, in part, by the DOE grant DE-SC0011842.

\end{document}